# T and CPT SYMMETRIES in ENTANGLED NEUTRAL MESON SYSTEMS

JOSE BERNABEU

Department of Theoretical Physics, University of Valencia, and
IFIC, Univ. Valencia-CSIC, E-46100 Burjassot, Valencia, Spain

Jose.bernabeu@uv.es

**Abstract**. Genuine tests of an asymmetry under T and/or CPT transformations imply the interchange between in-states and out-states. I explain a methodology to perform model-independent separate measurements of the three CP, T and CPT symmetry violations for transitions involving the decay of the neutral meson systems in B- and Φ-factories. It makes use of the quantum-mechanical entanglement only, for which the individual state of each neutral meson is not defined before the decay of its orthogonal partner. The final proof of the independence of the three asymmetries is that no other theoretical ingredient is involved and that the event sample corresponding to each case is different from the other two. The experimental analysis for the measurements of these three asymmetries as function of the time interval $\Delta t > 0$ between the first and second decays is discussed, as well as the significance of the expected results. In particular, one may advance a first observation of true, direct, evidence of Time-Reversal-Violation in B-factories by many standard deviations from zero, without any reference to, and independent of, CP-Violation.

In some quantum gravity framework the CPT-transformation is ill-defined, so there is a resulting loss of particle-antiparticle identity. This mechanism induces a breaking of the EPR correlation in the entanglement imposed by Bose statistics to the neutral meson system, the so-called ω-effect. I present results and prospects for the ω-parameter in the correlated neutral meson-antimeson states.

## 1. Introduction

I was asked to talk about T- and CPT-symmetries in the fundamental laws of Physics. The main point of these studies is that a genuine test of their invariance needs an interchange between in-states and out-states for a given process, a request particularly difficult to be accomplished for particles that decay during their time evolution. Can true Time Reversal Violation (TRV) be searched for in unstable systems? In this presentation I will advocate for a methodology in the neutral meson systems that makes use of the EPR-entanglement existing in B- and Φ- factories: in them, the preparation of a quantum mechanical individual state of the neutral meson is not made by measurements performed on it, but by the observation of the decay of its orthogonal correlated partner. This strategy will allow the quantum preparation of a given individual state of the (still living) neutral meson by selecting a particular decay channel of the other neutral meson.

Violation of CP invariance has been observed in the $K^0 - \overline{K}^0$ and $B^0 - \overline{B}^0$ systems. Up to now, the experimental results are in agreement [1] with the Standard Cabibbo-Kobayashi-Maskawa (CKM) mechanism in the ElectroWeak Theory. Although all present tests of CPT invariance confirm the validity of this symmetry, as imposed by any local quantum field theory with Lorentz invariance and Hermiticity [2], it would be of great interest to observe Time Reversal Violation (TRV) directly in a single experiment, independent on the question of CPT invariance. A direct evidence for true TRV would mean [3] an experiment that, considered by itself, clearly demonstrates T violation independent of, and unconnected to, the results for CP violation. There is at present no existing result that clearly shows TRV in this sense.

We are interested in Microscopic T-Symmetry Violations. Effects in particle physics odd under the change of the sign of time t ↔ -t are not necessarily T-violating. These observables can occur in theories with exact T-symmetry and are called T-odd effects, like those induced by absorptive components of the amplitude. Well known time asymmetries are the Universe t-asymmetry and the macroscopic t-asymmetry called the "arrow of time". But none of these t-asymmetries is a test of TRV. In the fundamental laws of physics, T-Violation exists in the Standard Model or any field theoretic extension of it. The observed CP-Violation in the neutral meson systems tells us that T should be violated as well. However, as emphasized above, true TRV has not been observed up to now. There are intriguing subtleties introduced by the antiunitary character of the symmetry operator and T-Violation means a non-vanishing asymmetry under the interchange of in ↔ out states.

There is no doubt that the Universe is expanding, even accelerating at present cosmological age. This natural t-asymmetry t ↔ -t is perfectly compatible with fundamental laws of physics that are Time Reversal-symmetric. It is due to the initial condition for our Universe, like Inflation. This asymmetry is similar to the fact that in our Universe we have a privileged reference frame, the one associated with the Cosmic Microwave Background (CMB) radiation at a definite temperature with fluctuations. In Figure 1 we show the WMAP result that allows fundamental measurements of cosmology, producing our New Standard Model of Cosmology.

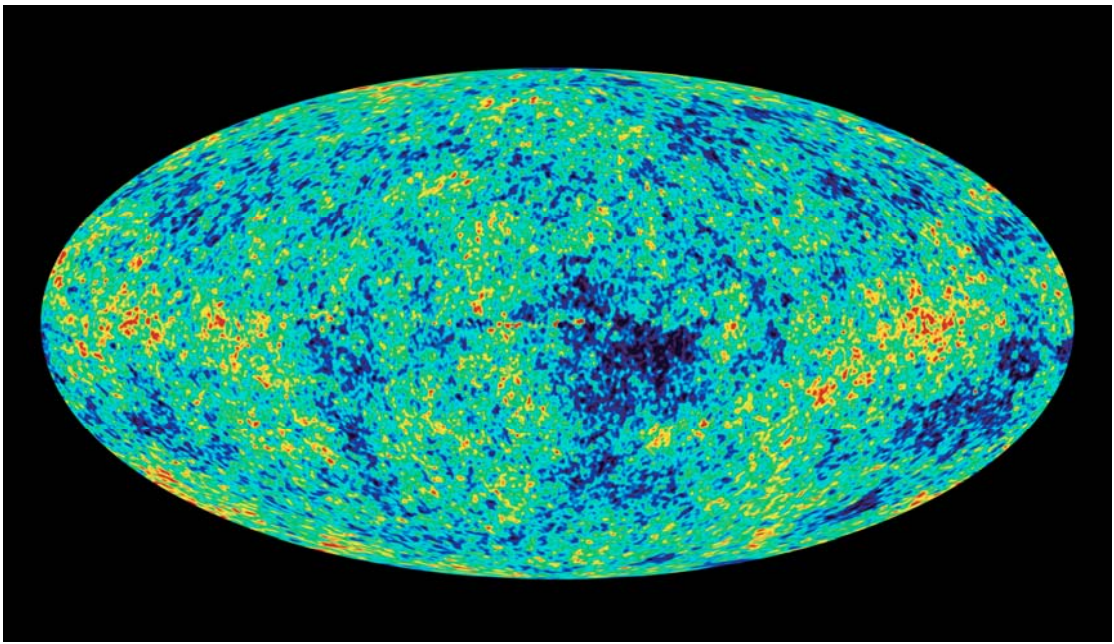

**Figure 1.** WMAP image of the CMB radiation

The CMB radiation has a thermal black body spectrum at a temperature of 2.725 K, as shown in Figure 2.

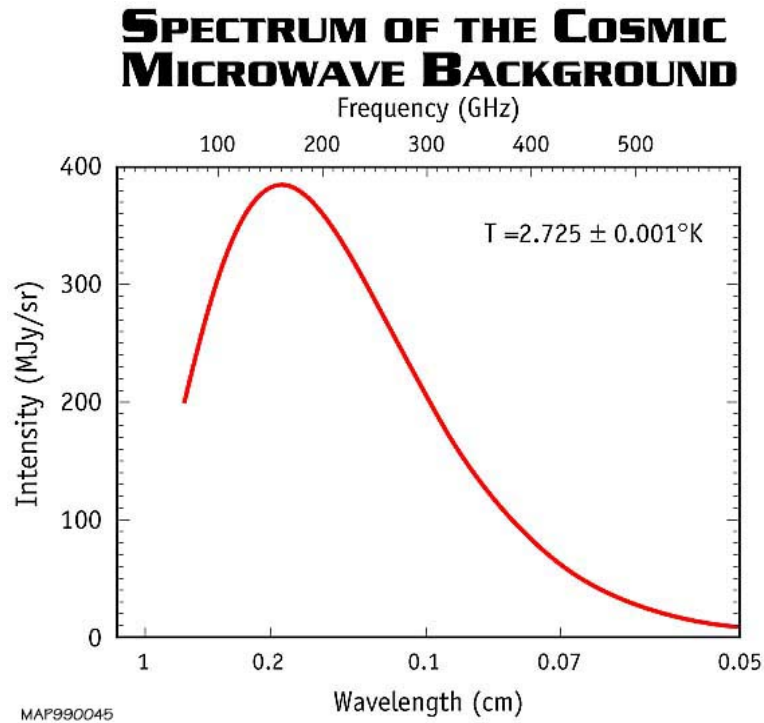

**Figure 2.** The CMB spectrum is the most precisely measured black body spectrum in nature

The spectrum peaks in the microwave range frequency of 160.2 BHz, corresponding to a 1.9 mm wavelength, in the intensity per unit frequency. This privileged reference frame in our Universe does not mean a violation of Lorentz invariance in the fundamental laws of physics. Similarly for the Universe time-asymmetry: it does not mean a violation of Time Reversal invariance in the fundamental laws of physics.

There is a macroscopic t-asymmetry, exemplified by the evolution of the Roman Coliseum in Figure 3, known as the "arrow of time".

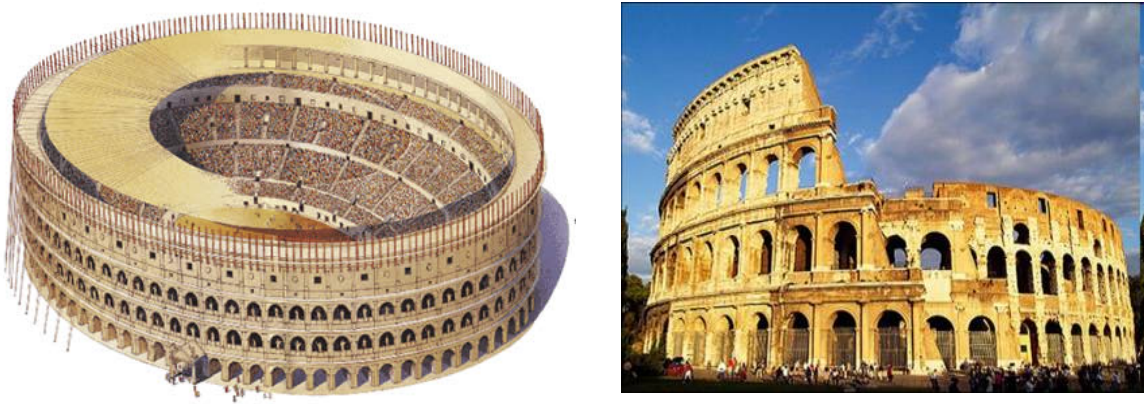

**Figura 3**. The "arrow of time" in the Roman Coliseum

This asymmetry is in the nature of Thermodynamics. According to Eddington [4], the Time's Arrow is a property of Entropy alone: Time is asymmetric with respect to the amount of order in an isolated system. A very interesting unsolved problem in quantum epistemology is the reduction of the wave packet: Is the quantum wave function collapse related to the thermodynamic arrow of time?

In particle physics, particle decays are an example of a time-asymmetric phenomenon. If we start with an initial collection of identical unstable particles we arrive after decay at a large collection of final states. There is little, if any, chance of any collection of such final states evolving to become a set of identical unstable particle as in the initial state. This phenomenon has nothing to do with T-violation. In fact, it looks like it prevents a true test of T-symmetry in unstable systems [3, 5], test that needs an exchange between "in" and "out" states.

In Section 2 we discuss the types of experiments that can provide a direct evidence for true microscopic T-violation and the method to circumvent the problem for unstable systems. In Section 3 we identify genuine observables for TRV, not needing $\Delta\Gamma \neq 0$, in the time evolution of the system. The corresponding strategy, based on the EPR-entanglement, is applied to CP, T and CPT symmetries as independent "experimental" transformations. Section 4 presents a Monte Carlo study for TRV and the experimental significance expected, from a $\chi^2$ test, for the TRV Asymmetries. In Section 5 we discuss the possible modifications in the EPR-entanglement induced by the ω-effect. Section 6 summarizes our conclusions.

## 2. Can T-symmetry be tested for unstable systems?

As emphasized before, a direct evidence for TRV would mean an experiment that, considered by itself, clearly shows TRV independent of, and unconnected to, the results for CP-violation (CPV). This evidence can be obtained from two types of experiments:

i) A non-zero expectation value of a T-odd operator for a non-degenerate stationary state, such as a non-vanishing Electric Dipole Moment (EDM), which is a P-odd, C-even, T-odd operator. The present experimental status in reviewed in [6] in this Proceedings.

The EDM can be generated by either Strong T-violation, like the θ-term $\epsilon_{\mu\nu\zeta\sigma} F^{\mu\nu} F^{\zeta\sigma}$ in the QCD Lagrangian, or Weak T-violation. The experimental small value of the θ-term needs a mechanism that protects it, the Peccei-Quinn symmetry [7]. The Standard Model theory of Electroweak Interactions or any local field theory with CP-violation, predicts T-violation effects in parallel.

ii) For a transition i→f, under the exchange in ↔ out, T-symmetry implies the connection $S_{f,i} \rightarrow S_{-i,-f}$, where –i(f) means the T-transformed state of the i(f) state.

The Kabir asymmetry $K^0 - \overline{K}^0$ vs. $\overline{K}^0 - K^0$ has been measured in CP-LEAR [8] with non-vanishing value. But $K^0 - \overline{K}^0$ is a CPT-even transition, so CP and T are experimentally identical here: a non-vanishing identically connected CP and T asymmetry needs $\Delta\Gamma \neq 0$ and the effect is time-independent.

The main question is: Is it possible to search for TRV in Mixing x Decay transitions? The origin here comes from the interference of decay amplitudes with and without mixing. We concentrate on Neutral Meson Factories. The opportunity arises [9, 10] in these Facilities from the quantum mechanical entanglement imposed by the Einstein-Podolsky-Rosen (EPR) correlation [11]. This correlation allows the quantum preparation of a physical state of a living particle by observing the decay of its partner particle: the individual state of each particle in the system is not defined before this observation. Depending on the selection of the decay channel, one can have separate tests of CP, T and CPT-symmetries!

In meson factories, the coherence between orthogonal $B^0, \overline{B}^0$ states has been used for flavour tag

$$|i\rangle = \frac{1}{\sqrt{2}}\left[B^0(t_1)\overline{B}^0(t_2) - \overline{B}^0(t_1)B^0(t_2)\right] \qquad (1)$$

where the states "1" and "2" are defined by the time of their decay with $t_1 < t_2$. The observation of $B^0 \rightarrow l^+$, for example, at time $t_1$, tells us that the complementary (still living) state is $\overline{B}^0$ at $t_1$. This is the preparation of the initial state for single state time evolution.

As said, the individual state of each neutral meson is, however, not defined before its collapse as a filter imposed by the observation of the decay of its companion. One can rewrite the same state (1) of the system $|i\rangle$ in terms of any other pair of orthogonal states of the individual neutral B-mesons: Considerer $B_+$ and $B_-$ where $B_-$ is the neutral B-state filtered by its decay to J/ψ $K_+$, $K_+$ being the neutral K-state filtered by its decay to π π, and $B_+$ is orthogonal to $B_-$, not connected to J/ψ $K_+$. As the final states are CP-eigenstates, we may call the filter imposed by a first observation, at time $t_1$, of one of these decays a "CP-tag" [12], although $B_\pm$ are not CP-eigenstates of B's necessarily.

The same entangled state of the two-body system can be rewritten

$$|i\rangle = \frac{1}{\sqrt{2}} [ B_+(t_1)B_-(t_2) - B_-(t_1)B_+(t_2) ] \qquad (2)$$

Now we may proceed to a partition of the complete set of events into four categories, defined by the tag in the first decay at $t_1$: $B_+, B_-, B^0 \text{ or } \overline{B}^0$, so we have eight different Decay-Intensities at our disposal, as functions of $\Delta t = t_2 - t_1 > 0$. Each one of these eight processes has an Intensity

$$I_i(\Delta t) \sim e^{\Gamma \Delta t} \{ C_i \cos(\Delta m \Delta t) + S_i \sin(\Delta m \Delta t) + C_i' \cosh(\Delta \Gamma \Delta t) + S_i' \sinh(\Delta \Gamma \Delta t) \} \qquad (3)$$

where Γ is the average width.

Up to now, for CPV analyses in B-factories, BABAR & BELLE have assumed CPT invariance and ΔΓ=0. In this case, there is a theorem which is operating [13]: Then Δt ↔ -Δt exchange, i.e., the exchange of the two decay products at $t_1$ and $t_2$, which is <u>not</u> a T-symmetry operation, becomes equivalent to T, i.,e., the exchange of the "in" and "out" neutral B-states. In this case, only two independent Intensities remain to be compared, when CP ~ T ~ Δt are theoretically connected through their equivalence.

We notice that the Intensities (3) contain terms independent of ΔΓ, in such a way that we will find asymmetries that, contrary to Kabir's asymmetry, do not need a non-vanishing ΔΓ associated to the decay properties.

### 3. Genuine observables not needing ΔΓ

We are now in the position of defining the genuine observables for testing the three symmetries CP, T and CPT <u>separately</u>.

1) Take $B_0 - B_+$ as the Reference transition and call (X, Y) the observed decay products at times $t_1$ and $t_2$, respectively. The CP, T and CPT transformed transitions are given in Table 1.

**Table 1.** The $B^0 \rightarrow B_+$ reference process and its CP, T and CPT transformed processes.

| **Transition** | $B^0 \rightarrow B_+$ | $\overline{B}^0 \rightarrow B_+$ | $B_+ \rightarrow B^0$ | $B_+ \rightarrow \overline{B}^0$ |
|---|---|---|---|---|
| (X,Y) | (l⁻,ΨK$_L$) | (l⁺,ΨK$_L$) | (J/ΨK$_s$, l⁺) | (J/ΨK$_s$, l⁻) |
| Transformation | Reference | CP | T | CPT |

The four processes are experimentally independent, as you may check, and the Δt-exchanged processes X↔Y are <u>not</u> in Table 1. This last comment is particularly important, because (X, Y) and (Y, X) are usually included in the same experimental "sample".

2) Select (Y, X) from 1) as Reference, associated with $B_- \to \overline{B}_0$ transition, and consider its three CP, T and CPT transformed transitions. We have thus four additional processes which are experimentally independent among themselves and with those considered in Table 1. These eight independent transitions are described by the Decay-Intensities given in equation (3).

We thus conclude that 6 independent Asymmetries, 2 for each CP, T and CPT transformations, plus the 2 Intensities for the Reference processes can be built.

The 2 genuine CPV-Asymmetries are

$$\Delta t \left[ \begin{array}{l} A_{CP,1} = \dfrac{\Gamma(l^-, J/\psi\, K_L) - \Gamma(l^+, J/\psi\, K_L)}{\Gamma(l^-, J/\psi\, K_L) + \Gamma(l^+, J/\psi\, K_L)} \\ \\ A_{CP,2} = \dfrac{\Gamma(J/\psi\, K_L, l^-) - \Gamma(J/\psi\, K_L, l^+)}{\Gamma(J/\psi\, K_L, l^-) + \Gamma(J/\psi\, K_L, l^+)} \end{array} \right. \qquad (4)$$

The 2 genuine TRV-Asymmetries are

$$\Delta t \left[ \begin{array}{l} A_{T,1} = \dfrac{\Gamma(l^-, J/\psi\, K_L) - \Gamma(J/\psi\, K_S, l^+)}{\Gamma(l^-, J/\psi\, K_L) + \Gamma(J/\psi\, K_S, l^+)} \\ \\ A_{T,2} = \dfrac{\Gamma(J/\psi\, K_L, l^-) - \Gamma(l^+, J/\psi\, K_S)}{\Gamma(J/\psi\, K_L, l^-) + \Gamma(l^+, J/\psi\, K_S)} \end{array} \right. \qquad (5)$$

The 2 genuine CPTV-Asymmetries are

$$\Delta t \left[ \begin{array}{l} A_{CPT,1} = \dfrac{\Gamma(l^-, J/\psi\, K_L) - \Gamma(J/\psi\, K_S, l^-)}{\Gamma(l^-, J/\psi\, K_L) + \Gamma(J/\psi\, K_S, l^-)} \\ \\ A_{CPT,2} = \dfrac{\Gamma(J/\psi\, K_L, l^-) - \Gamma(l^-, J/\psi\, K_S)}{\Gamma(J/\psi\, K_L, l^-) + \Gamma(l^-, J/\psi\, K_S)} \end{array} \right. \qquad (6)$$

As noticed above, the 6 Asymmetries (4), (5) and (6) are <u>independent</u>, no matter that the CPT-operation is the product of CP and T transformation. However, for a given symmetry, one can increase the statistical significance of the experimental study by considering additional processes taken as a Reference. Thus we have

3) Take $B^0 \to B_-$ as the Reference transition. The CP, T and CPT transformed transitions are

**Table 2.** The $B^0 \to B_-$ reference process and its CP, T and CPT transformed processes.

| Transition | $B^0 \to B_-$ | $\overline{B}^0 \to B_-$ | $B_- \to B^0$ | $B_- \to \overline{B}^0$ |
|---|---|---|---|---|
| (X,Y) | (l⁻,J/ΨK$_S$) | (l⁺,J/ΨK$_S$) | (J/ΨK$_L$, l⁺) | (J/ΨK$_L$, l⁻) |
| Transformation | Reference | CP | T | CPT |

Although these four processes are the same as those considered in 2), they can be combined in different ways for each of the three symmetries. For example, the CP-Asymmetry in 2) [see equation (4)] involves the decay product to J/ψ K$_L$ only. In Table 2, however, referring to the choice of reference 3), one obtains a CP-Asymmetry involving J/ψ K$_S$ instead. Another example: the CPT-Asymmetry in 2) [see equation (6)] involves the decay product to l⁻ only. In Table 2, however, one can obtain a CPT-Asymmetry involving the decay product to l⁺ instead, changing the Reference.

4) Select (Y, X) from 3) as Reference, associated with $B_+ \to \overline{B}^0$ transition, and consider its three CP, T and CPT transformed transitions. The four processes are experimentally independent among themselves and with those considered in 3), i.e., in Table 2. Although these four processes in 4) are the same as those considered in 1), they can be combined in different ways for each of the three symmetries.

We thus conclude that there are another 6 Asymmetries, 2 additional ones for each CP, T and CPT transformations.

The 2 additional genuine CPV-Asymmetries are

$$\Delta t \left[ \begin{array}{l} A_{CP,3} = \dfrac{\Gamma(l^-, J/\psi K_S) - \Gamma(l^+, J/\psi K_S)}{\Gamma(l^-, J/\psi K_S) + \Gamma(l^+, J/\psi K_S)} \\ \\ A_{CP,4} = \dfrac{\Gamma(J/\psi K_S, l^-) - \Gamma(J/\psi K_S, l^+)}{\Gamma(J/\psi K_S, l^-) + \Gamma(J/\psi K_S, l^+)} \end{array} \right. \quad (7)$$

The 2 additional genuine TRV-Asymmetries are

$$\Delta t \left[ \begin{array}{l} A_{T,3} = \dfrac{\Gamma(l^-, J/\psi K_S) - \Gamma(J/\psi K_L, l^+)}{\Gamma(l^-, J/\psi K_S) + \Gamma(J/\psi K_L, l^+)} \\ \\ A_{T,4} = \dfrac{\Gamma(J/\psi K_S, l^-) - \Gamma(l^+, J/\psi K_L)}{\Gamma(J/\psi K_S, l^-) + \Gamma(l^+, J/\psi K_L)} \end{array} \right. \quad (8)$$

The 2 additional genuine CPTV-Asymmetries are

$$\Delta t \begin{cases} A_{CPT,3} = \dfrac{\Gamma(l^+, J/\psi\, K_S) - \Gamma(J/\psi\, K_L, l^+)}{\Gamma(l^+, J/\psi\, K_S) + \Gamma(J/\psi\, K_L, l^+)} \\[2ex] A_{CPT,4} = \dfrac{\Gamma(J/\psi\, K_S, l^+) - \Gamma(l^+, J/\psi\, K_L)}{\Gamma(J/\psi\, K_S, l^+) + \Gamma(l^+, J/\psi\, K_L)} \end{cases} \quad (9)$$

For a CPV test, we should use (4) plus (7) asymmetries. For a TRV test, we should use (5) plus (8) asymmetries. They are model-independent in the sense that this experimental analysis is free from any theoretical prejudice: only the EPR-correlation has been used to prepare the state of the individual neutral B. The experimental channels involved in each symmetry test are different.

And, what for genuine Asymmetries in DAPHNE ? For $K^0 \to \overline{K}^0$, contrary to $B^0 \to \overline{B}^0$, the "physical" states of definite mass and width have $\Delta\Gamma \neq 0$, so that the theorem [13] discussed before for CPT invariance, i.e., the equivalence of T (and CP) with $\Delta t$-exchange, is no longer applicable.

If $K_+$ is the neutral K-state filtered by the decay to $2\pi$ and $K_-$ is its orthogonal state, the Master Table for the Φ-factory becomes

**Table 3**. Four asymmetries for each CP, T, CPT transformations in DAPHNE

| Reference | CP | T | CPT |
|---|---|---|---|
| $K^0 \to K_+$ $(l^-, 2\pi)$ | $\overline{K}^0 \to K_+$ $(l^+, 2\pi)$ | $K_+ \to K_0$ $(3\pi, l^+)$ | $K_+ \to \overline{K}^0$ $(3\pi, l^-)$ |
| $K_- \to \overline{K}^0$ $(2\pi, l^-)$ | $K_- \to K^0$ $(2\pi, l^+)$ | $\overline{K}^0 \to K_-$ $(l^+, 3\pi)$ | $K^0 \to K_-$ $(l^-, 3\pi)$ |
| $K^0 \to K_-$ $(l^-, 3\pi)$ | $\overline{K}^0 \to K_-$ $(l^+, 3\pi)$ | $K_- \to K_0$ $(2\pi, l^+)$ | $K_- \to \overline{K}^0$ $(2\pi, l^-)$ |
| $K_+ \to \overline{K}^0$ $(3\pi, l^-)$ | $K_+ \to K^0$ $(3\pi, l^+)$ | $\overline{K}^0 \to K_+$ $(l^+, 2\pi)$ | $K^0 \to K_+$ $(l^-, 2\pi)$ |

Experimental analyses of CP, T, CPT Asymmetries with DAΦNE data are being pursued [14]

## 4. Monte Carlo study for TRV in the B-system

We concentrate here on the four asymmetries for true TRV in the neutral B-system, i.e., equations (5) and (8): $B^0 \leftrightarrow B_+$, $B_- \leftrightarrow \bar{B}^0$, $B^0 \leftrightarrow B_-$, $B_+ \leftrightarrow \bar{B}^0$ as Reference.

A PDF allowing CP, T & CPT violation parameters has been developed in [15] for generating events. The true value has been taken as in the Standard Model: CPT-symmetry, $\Delta\Gamma = 0$, $C_i = 0$, $S_i = \pm 0.672$ (sin 2 β) in the Intensities of equation (3). For the $A_T$ asymmetries, as function of Δt, Figure 4 gives the results including proper-time resolution, mistags and efficiency effects (as taken from BABAR published papers):

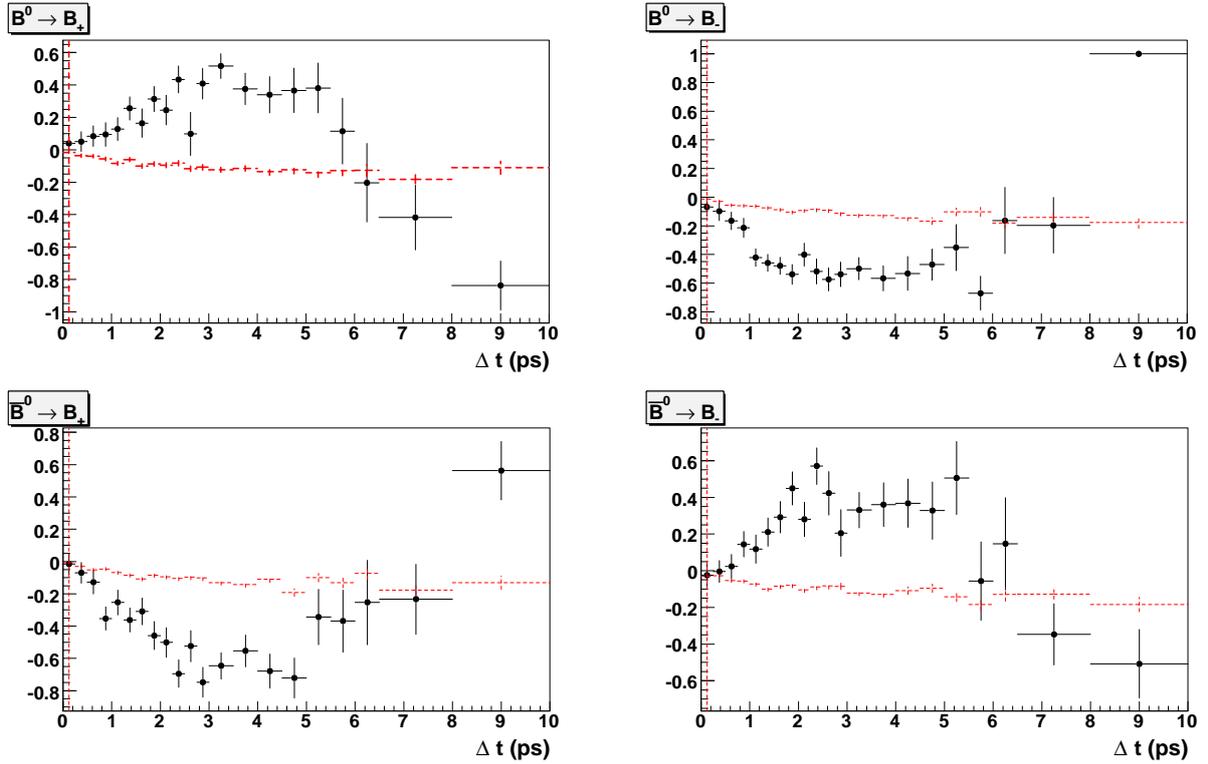

**Figure 4**. The four TRV-Asymmetries associated with $B_0 \leftrightarrow B_+$, $B_- \leftrightarrow \bar{B}^0$, $B_0 \leftrightarrow B_-$, $B_+ \leftrightarrow \bar{B}^0$. The red points give the expected experimental asymmetry for absence of true TRV.

Each one of the four $A_T$-Asymmetries has an experimental significance obtained from a $\chi^2$ test. The main conclusion, as seen in Table 4, is that one gets much more than a 5σ-effect!

**Table 4**. Experimental significance for each of the 4 TRV-Asymmetries.

| TRV test | $B^0 \to B_+$ | $B^0 \to B_-$ | $\bar{B}^0 \to B_+$ | $\bar{B}^0 \to B_-$ |
|---|---|---|---|---|
| Standard Deviations | 6.70 | 9.84 | 9.42 | 7.34 |

As the four results are statistically independent, they could be combined in the $\chi^2$-test for obtaining a global significance, assuming that the four asymmetries are theoretically related.

These simulated results provide an expected "guarantee" of a significant discovery for true TRV. A fundamental experimental result would be a first observation of true, direct evidence for genuine TRV by many standard deviations from zero, without any reference to, and independent of, CPV.

Working with the four CP-Asymmetries of Equations (4) and (7), one expects similar levels of experimental significances independently.

## 5. The ω-effect

The proposed tests of separate CP, T, CPT symmetries in the neutral meson systems are based on the EPR-Entanglement existing in the Meson Factories as a consequence of Particle Identity: $K^0, \overline{K}^0$ are two states of identical particles, connected by CPT.

Besides the permutation operation $\mathscr{P}$ for space-time properties, the strangeness charge connection is made by C, so that for bosons the indistinguishibility requirement is C $\mathscr{P}$ = +.

In neutral meson factories, $K^0, \overline{K}^0$ are produced by Φ-decay with J=1, S=0. This implies L=1 and C= -, so that $\mathscr{P}$ = -, i.e., an <u>antisymmetric wave function</u>. This antisymmetry is responsible for preserving $K^0 \overline{K}^0$ terms only in the time evolution of the two-body system, including the Mixing $K^0 \leftrightarrow \overline{K}^0$. Similarly for $K_+ K_-$ terms only. This correlation is perfect for tagging: Flavour-Tag, CP-Tag,…

The question is [16]: What if the $K^0 \overline{K}^0$ Identity is lost? In this case, the two particle system would not satisfy the requirement C$\mathscr{P}$ = +. In perturbation theory, if still J=1 with C=-, this breaking leads to a mixing of the "forbidden" $\mathscr{P}$ = + symmetric state into the "allowed" $\mathscr{P}$ = - antisymmetric state:

$$|i\rangle = |antisymmetric\rangle + \omega |symmetric\rangle \qquad (10)$$

This perturbative mixing is the ω-effect: In the time evolution of the system one finds now ω $K_0 K_0$, ω $K_+ K_+$ terms,…, i.e., a Demise of Tagging.

In some Quantum Gravity models, matter propagation in topologically non-trivial space-time vacua suffers a possible loss of quantum coherence or "decoherence". This effect can be originated by space-time foam backgrounds [17]. The matter quantum system is an open system, interacting with the "environment" of quantum gravitational degrees of freedom: this interaction leads to an apparent loss of unitarity for low-energy observers. As a consequence, there is not a well-defined S-matrix between asymptotic states and the CPT operator is not well defined [18].

This kind of CPT-non definition should be disentangled from the case of effective theories for Lorentz invariance violation [19], in which CPT-breaking means [$H_{eff}$, CPT] ≠ 0. The CPT-Violation discussed here would be an "intrinsic" microscopic time irreversibility, so that $\overline{K}^0$ is not "well-defined" from $K^0$. It implies: 1) a modified single $K^0, \overline{K}^0$ evolution, leading to the (α, β, γ) parameterization [20]; 2) for entangled Kaon states in a Φ-factory, the ω-effect.

Looking for observables which are a signal of the ω-effect, consider [16] the Φ-decay amplitude described by Figure 5 and equation (11):

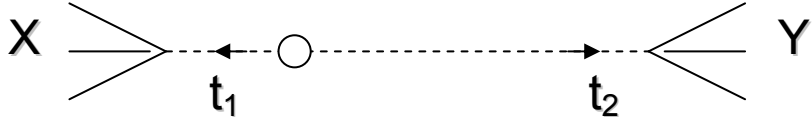

**Figure 5**. $K^0 - \overline{K}^0$ decay to (X, Y) at times $t_1$ and $t_2$, respectively

$$A(X,Y) = \langle X|K_S \rangle \langle Y|K_S \rangle N(A_1 + A_2)$$
$$A_1 = e^{-i(\lambda_L + \lambda_S)t/2}\left[\eta_X e^{-i\Delta\lambda\Delta t/2} - \eta_Y e^{i\Delta\lambda\Delta t/2}\right] \quad (11)$$
$$A_2 = \omega\left[e^{-i\lambda_S t} - \eta_X \eta_Y e^{-i\lambda_L t}\right]$$

where $A_1$ is the "allowed" amplitude, whereas $A_2$ is the "forbidden" amplitude proportional to $\omega$.

An inspection of equation (11) tells us that the good strategy to enhance the relative effect of $\omega$ is the selection of a decay channel suppressed by the $\eta$'s, the ratio between the $K_L$ and $K_S$ decay amplitudes. This requirement is fulfilled by the choice $X = Y = \pi^+\pi^-$, a CP "forbidden" channel in K-physics, with $\eta$ a small ratio. The relative $\omega$-effect in equation (11) is thus $\omega/|\eta_{+-}|$.

From equation (11) we can calculate the Intensity as function of $\Delta t$

$$I(\Delta t) = \frac{1}{2}\int_{\Delta t}^{\infty} dt |A(X,Y)|^2 \quad (12)$$

For the ($\pi^+\pi^-, \pi^+\pi^-$) channel, and writing $\omega = |\omega|e^{i\Omega}, \eta_{+-} = |\eta_{+-}|e^{i\Phi_{+-}}$, the Intensity (12) is plotted in Figure 6 for $\omega = 0$ (continuous line) and an $\omega$-value comparable to $\eta_{+-}$ (broken line).

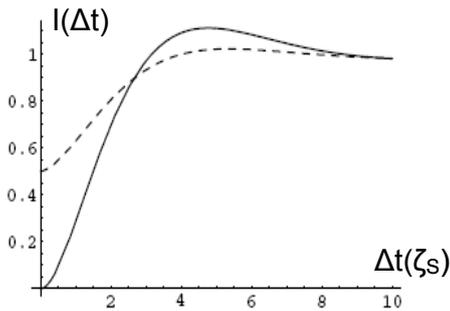

**Figure 6**. The Intensity $I(\Delta t)$ for the decay of $K^0 - \overline{K}^0$ to ($\pi^+\pi^-, \pi^+\pi^-$), for $\omega=0$ ( —— ) and $|\omega| = |\eta_{+-}|, \Omega = \Phi_{+-} - 0.16\pi$ (---)

The most prominent effect is the breaking of $I(\Delta t) \sim 0$ for small values of $\Delta t$, a result that was a consequence of the particle identity anti-correlation: no identical states at $t_1 = t_2$.

The KLOE experiment has obtained the first measurement of the $\omega$-parameter [21]:

$$\left.\begin{array}{l}\text{Re}(\omega) = \left(-1.6^{+3.0}_{-2.1\,stat} \pm 0.4_{syst}\right)\times 10^{-4}\\ \text{Im}(\omega) = \left(-1.7^{+3.3}_{-3.0\,stat} \pm 1.2_{syst}\right)\times 10^{-4}\end{array}\right\} \quad |\omega| < 1.0 \times 10^{-3} \text{ at } 95\% \text{ CL} \quad (13)$$

At least one order of magnitude improvement is expected with KLOE-2 at the upgraded DA$\Phi$NE.

All decoherence effects, including the ω-effect, manifest as a deviation from the quantum mechanical prediction of the EPR-correlation I($\pi^+ \pi^-, \pi^+ \pi^-$; $\Delta t = 0$) = 0. Hence the reconstruction of events in the region near $\Delta t \sim 0$ is crucial. Experimentally, one needs vertex resolution. In Figure 7 one finds a Monte Carlo simulation of I($\pi^+ \pi^-, \pi^+ \pi^-$; $\Delta t$), with both the KLOE resolution $\sigma_{\Delta t} \approx \zeta_S$ (black histogram) and the expected KLOE-2 resolution $\sigma_{\Delta t} \approx 0.3 \zeta_S$ (red histogram)

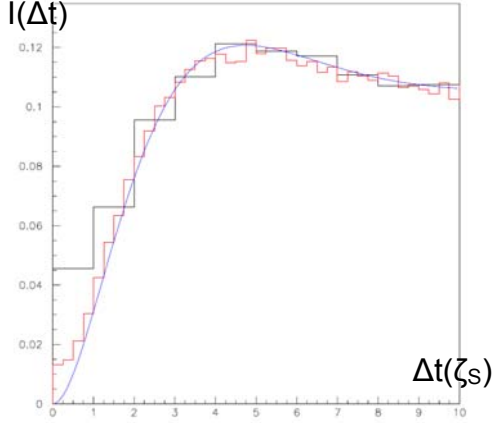

**Figure 7**. Comparision of the I ($\pi^+ \pi^-, \pi^+ \pi^-$) simulation for KLOE $\Delta t$ resolution ( —— ) and KLOE-2 $\Delta t$-resolution ( —— )

In B-factories, there is no privileged channel to enhance the ω-effect. With currently available data from BABAR and BELLE, the CP-violating semileptonic charge asymmetry, in the equal sign dilepton channels I ($l^\pm, l^\pm$; $\Delta t$) gives the bounds [22]

$$-0.0084 \leq \text{Re}(\omega) \leq 0.0100 \text{ at } 95\%\text{CL} \qquad (14)$$

The ω-effect has been described as the result of local distortions of space-time in the neighborhood of defects, interacting with matter [23]. The recoil of the Planck-mass defect leads to a metric deformation

$$g_{0i} \sim \Delta k^i / M_P = \zeta k^i / M_P \qquad (15)$$

In this framework, Lorentz invariance still holds macroscopically $< \zeta k^i > = 0$, but one has non-trivial quantum fluctuations $< \zeta^2 k^i k^j > \sim \delta_{ij} \zeta^2 | \vec{k} |^2$. The stochastic effects of the space-time foam lead to an ω-value

$$|\omega|^2 \sim \frac{\zeta^2 |\vec{k}|^4}{M_P^2 \Delta m^2} \qquad (16)$$

which is enhanced by the quasi-degeneracy of the mass eigenstates.

At the DAΦNE energy, equation (16) gives $|\omega| \sim 10^{-4} \zeta$, which lies within the sensitivity of KLOE-2 for not much small values of the momentum transfer fraction $\zeta$.

## 6. Conclusions
The observed time-Asymmetries in the Universe and the macroscopic "arrow of time" are not Time-Reversal Violating. A true fundamental TRV means an Asymmetry under the exchange of in ↔ out states.

For unstable systems, there is unique opportunity for preparing the quantum states needed in a true T-symmetry test by using the EPR-Entanglement between the two neutral mesons in B and Φ factories. The Golden Channels associated with Mixing x Decay amplitudes offer 8 different Decay-Intensities.

For each one of the three symmetries CP, T and CPT, one can work with appropriate combinations of the Intensities for generating 4 Genuine Independent Asymmetries.

A Monte Carlo study for true TRV in B-factories leads to simulated results for each of the four asymmetries with more than 5 standard deviations from zero. We have thus an expectation of having experimental results leading to a first discovery of true TRV independent of, and unconnected to, CPV.

In certain quantum gravity models, the S-matrix and the CPT-operator are not well defined, leading to a breaking of the $K^0, \overline{K}^0$ particle identity and the appearance of the ω-effect. The measurement of the "forbidden" amplitude induced by the ω-parameter is better done with the privileged channel in the Φ-factory I($\pi^+\pi^-$, $\pi^+\pi^-$; $\Delta t$). A sensitivity to $|\omega| \sim 10^{-4}$ at KLOE-2 is not far from expectations in some models of space-time foam.

**Acknowledgments**
I would like to thank the conceptual discussions maintained with several colleagues, specially Antonio Di Domenico, Fernando Martinez and Pablo Villanueva. This work is supported by the Spanish Ministry MICINN and Generalitat Valenciana under Grants FPA-2008-02878 and PROMETEO-2008/004.